# Two-dimensional porous silicon photonic crystal light emitters

Maria Makarova, Jelena Vuckovic, Hiroyuki Sanda, Yoshio Nishi
*Department of Electrical Engineering, Stanford University, Stanford, CA 94305-4088*

**Abstract**

We present the design, fabrication and preliminary experiments of two-dimensional photonic crystal cavities made in nanoporous silicon luminescent at 700-800 nm. Enhancement in photoluminescence extraction efficiency at the resonant wavelength is expected due to Purcell effect and directed radiation pattern defined by the cavity. Such cavities should also enhance nonlinearities exhibited by porous Si beyond what is observed in one-dimensional distributed Bragg reflection cavities due to their small mode volumes and modest quality factors. This design aligns itself well to integration with conventional silicon based electronics on a single chip.

We propose to study light emitters based on two-dimensional planar photonic crystals (PCs) fabricated in nanoporous silicon (pSi) that exhibits luminescence. Porous silicon luminescent properties have been studied for decades[1]. Recently external quantum photoluminescence efficiency reaching 23% at room temperature was demonstrated as opposed to 1-2% observed typically[2]. Confining luminescent material in an optical micro-cavity would increase the efficiency further by restricting the resonant wavelength to a directed radiation pattern that can be collected effectively, and by reducing the radiative lifetime of the on-resonance emitters while suppressing off-resonance emission through Purcell effect[3,4]. The idea of using micro-cavities to improve pSi emission characteristics was first proposed and tested in the context of one-dimensional Bragg reflector cavities achieved by porosity modulation in the mirrors. Drastic PL bandwidth narrowing and 16-fold increase in the peak intensity were observed[5]. These results are most likely due to strong directionality of output, since Purcell effect is insignificant in such cavities[6,7]. We propose 2-D PC based micro-cavities for this task because of their high quality (Q) factors and small mode volumes (V) necessary for Purcell effect. Planar geometry of such implementation favors integration with other optical devices on chip. In addition, implementation of photonic crystals on Si and pSi platform leads to a possibility of integrating electronic and optical devices on the same chip, either monolithically or by device layer transfer. Here we present a design for making such photonic crystals cavities supported by numerical simulations, discuss fabrication parameters, and demonstrate preliminary results.

Figure 1(a) shows top and side views of proposed PC structure. Hexagonal array of holes is drilled in the luminescent pSi core layer sitting on top of a higher porosity and thus lower refractive index substrate layer made on a Si wafer. Lower refractive index in the substrate layer is necessary to provide vertical confinement for photonic crystal structure through total internal refraction. The substrate layer is made thick enough such that the underlying interface with bulk Si is not seen by the optical modes confined in the core. Accordingly the underlying bulk Si will be omitted from analysis. Assumed refractive indices, which depend on pSi fabrication as discussed later, are 2 for the luminescent core layer, and 1.4 for substrate. The main design parameters are the air hole radius **r**, drill depth **d**, core thickness **c** in relation to inter-hole spacing **a**.

To localize optical modes in a PC cavity it is necessary to first design PC that has a photonic bandgap. We used three-dimensional finite difference time-domain (3D FDTD) method[8] to simulate band structure of the PC with various parameters. A single unit cell with 20 computational points per inter-hole spacing *a* was simulated with periodic Bloch boundary conditions in-plane and absorbing MUR boundary conditions out of plane. Band gap of 16% extending from 0.4616 to 0.5415 $a/\lambda$ was achieved for TE-like modes with **r**=0.4*a*, **c** =0.75*a*, **d**=1.5*a* as shown in Fig. 1(b). It is worth noticing that drilling the holes of PC deep into the cladding layer was key to achieving a band gap with this relatively low index contrast between the core and the cladding[9]. Intuitively, drilling holes deeper makes the effective refractive index of the substrate lower and improves vertical confinement; in addition it makes the structure more symmetric vertically which helps to separate TE and TM modes. The area shaded in gray on the band diagram corresponds to all the modes that would leak into air from the core, and cannot be confined.

A linear cavity formed by eliminating 3 air holes along $\Gamma$J direction[10] was designed based on the PC described above. Figure 2 shows the cavity with the simulated field distribution patterns. Vertical cross-section of the structure with overplayed electric field in y direction shows that the field is primarily concentrated in the core photoluminescent material and extends only slightly into the substrate and air regions. The cavity quality factor is limited by the radiation that escapes vertically into air and substrate, since the in-plane confinement can be improved by increasing the size of photonic crystal mirrors around the cavity. The end holes of the cavity were shifted outward by 0.15**a** to increase the quality factor as suggested by reference 10. The cavity out-of-plane Q-factor, which limits the total Q-factor[8], predicted by the 3D FDTD simulation is 294 and mode volume is V=3.5 $(\lambda/n)^3$. The expected maximum Purcell factor, i.e., reduction of radiative lifetime inside the cavity relative to that in bulk porous Si is 6.4 as given by $3/(4\pi^2) (\lambda/n)^3$ Q/V [3,4]

Fabrication of photonic structures described so far is simple: first, pSi layers are defined by anodization, then PC is made in porous layers by a combination of e-beam lithography and dry etching. Both of these procedures are well established, but require some process tuning to achieve good fabrication results. In our simulation, the refractive index of the core layer is assumed to be 2 implying porosity of around 50%, while that of the substrate layer is assumed to be 1.4 implying porosity of 80-90%, which is possible by simply increasing current after the anodization of the core layer[11]. Generally porosity above 50% is required to achieve luminescence[1], but slight oxidation of pSi can be used to attain luminescence while keeping refractive index at 2 for the core layer. We demonstrate the feasibility of fabricating photonic crystal structures in a single thick layer of porous Si material luminescent around 700-750nm.

A (100) 4 inch p type silicon wafer with a resistivity of 0.7 cm was first anodized at a current density of 21mA/cm$^2$ for 5 minutes in the electrolyte consisting of 49% hydrofluoric acid, isopropanol, and deionized water in the volume ratio of 2:1:2. The thickness of anodized layer is approximately 1 μm at the center, while the silicon surface is etched to a depth of 4 to 20 μm with no pSi remaining around the edge probably due to larger current density at the edge. A piece from the center was used to make photonic crystal cavities. Fabrication of photonic crystal in pSi is similar to the fabrication of PCs in silicon on insulator, where PMMA layer is used both as an e-beam resist, and as a mask for dry etching[12]. First, PMMA with molecular weight of 495K and concentration of 5% in anisole was spun on the sample surface and baked on a hot plate at 150˚C for 15 min. Next, electron-beam lithography was performed in Raith 150 system at 10 keV and 10μ aperture to define photonic crystal cavity pattern. Subsequently, the sample was developed in 3:1 IPA:MIBK mixture for 50 sec. and rinsed in IPA for 30 sec., such that the exposed PMMA is removed. The pattern in PMMA was transferred into pSi layer by magnetically induced reactive ion etch with $CF_4$ and $HBr/Cl_2$ gas combinations. Figure 3a shows a SEM picture of the fabricated structure with **a**=400 nm and **r/a**=0.4.

Optical characterization setup used to measure photoluminescence from pSi and photonic crystal cavities is shown in Figure 4. 100X objective lens with NA=0.5, and working distance of 12.0 mm is used for collection of the luminescence and to illuminate the sample with white light for imaging. Green 5 mW pump laser operating at 532nm is focused on the sample from a side. The signal may be viewed on a CCD camera or collected by an optical fiber leading to a spectrometer. Iris placed in the focal plane may be used to spatially select region of interest for measurements. A high-pass filter at 600nm is used to filter out scattered laser radiation. Figure 5 shows the photoluminescence spectrum measured with this set up from a bulk porous Si sample. The peak is broad and is centered around 700 nm. Cavities fabricated in this luminescent material (such as the ones showin in Fig. 3a) are expected to narrow this peak around the design wavelength and increase its magnitude. Figure 3b shows a CCD image obtained with white light illumination, while figure 3c shows the photoluminescence emerging from this cavity under green laser excitation. Any scattered laser light was filtered, so we only see emission from the sample on the picture.

The PC forming the cavity shown here has periodicity of 400nm, which has its band gap above the luminescent region of the pSi sample, so we were not able to observe a strong modification in the emission spectrum, but the CCD image suggests that there is some interaction of emission with the guided bands of photonic crystal. We are in the process of fabricating cavities with smaller periodicities that will have band gap in the photoluminescent region, and would thus lead to an even stronger enhancement of photoluminescence.

In conclusion, we have proposed a design of a two-dimensional PC cavity with a small mode volume and a modest Q factor that utilized small index contrast of two pSi layers . Such cavities are expected to increase external photoluminescence efficiency of pSi through the Purcell effect enhancement and by restricting the radiation direction for easier collection. The cavities are also of interest for the study of nonlinear effect in porous Si[13,14]. Potential applications of the proposed structures are numerous. Efficient porous Si diodes should be possible to achieve with this scheme since electroluminescence of porous Si has already been demonstrated[15,16]. Integration of a light source on-chip with conventional Si microelectronics is possible. Photoluminescence of porous Si can be tuned over a wide range in the visible and NIR to be useful for a variety of devices. The current operation range around 750 nm is particularly interesting for biosensors to determine proportion of oxygenated hemoglobin.

This work has been supported in part by the CIS Seed Fund and MARCO Interconnect Focus Center.

**Figure Captions**

**Figure 1.**

(a) Top and side views of proposed photonic crystal structure consisting of hexagonal array of air holes with radius **r**, depth **d**, and inter-hole spacing **a** made on luminescent porous Si core layer with refractive index of 2 and thickness **c**, followed by porous Si substrate layer with refractive index of 1.4 and thickness **s** sitting on top of a Si wafer. The hexagon represents the first Brillouin zone with high symmetry points $\Gamma$, X, J indicated.

(b) Band diagram of TE-like modes for the structure with the following parameters: r/a=0.4, c/a=0.75, d/a=1.5, s/a>3. Gray region corresponds to the region above the light line in air.

**Figure 2.**

Electric and magnetic field patterns of the linear photonic crystal cavity formed by eliminating 3 air holes along $\Gamma$J direction and shifting the end holes out by 0.15a.

**Figure 3.**

(a) SEM image of the fabricated photonic crystal cavity with **a**=400 nm and **r/a**=0.4 in luminescent pSi.

(b) CCD camera white light image of the cavity.

(c) CCD camera image of the photoluminescence measured from the cavity.

**Figure 4.**

Schematic of optical setup used for photoluminescence measurements. Objective lens with NA 0.5, working distance of 12.0 mm, is used to illuminate the sample with white light for imaging, and for collection of the photoluminescence. Green 5 mW pump laser operating at 532nm is focused on the sample from a side. The signal may be collected by a CCD camera or an optical fiber leading to a spectrometer. Iris placed in the focal plane may be used to spatially select region of interest for measurements. A high-pass filter at 600nm is used to filter out scattered laser radiation.

**Figure 5.**

Photoluminescence from bulk pSi.

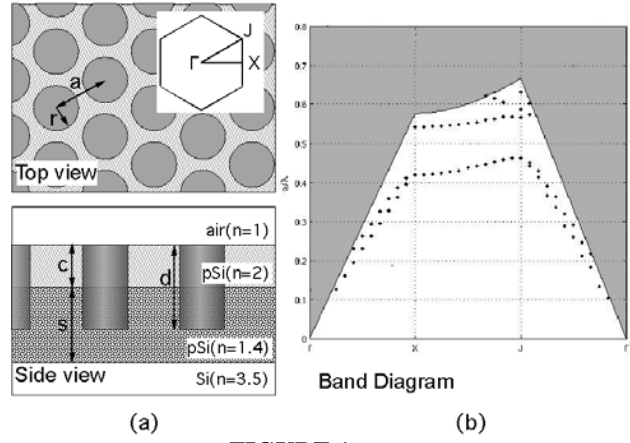

**FIGURE 1**

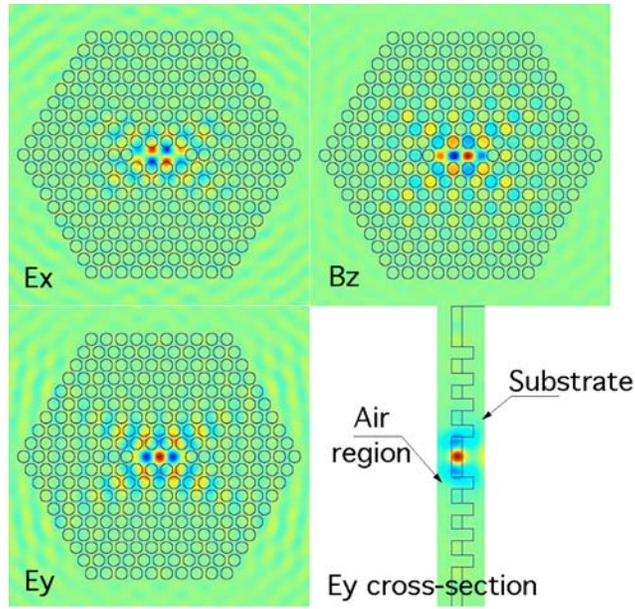

**FIGURE 2**

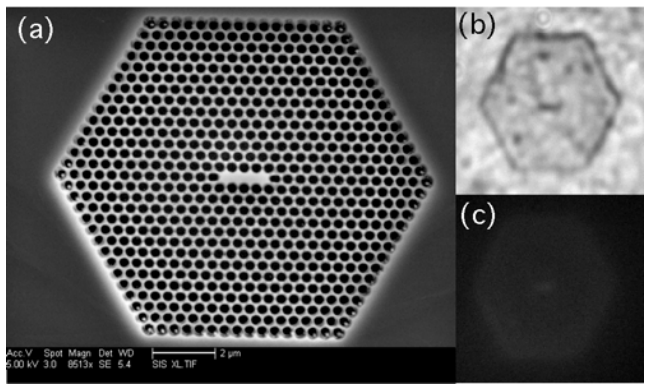

**FIGURE 3**

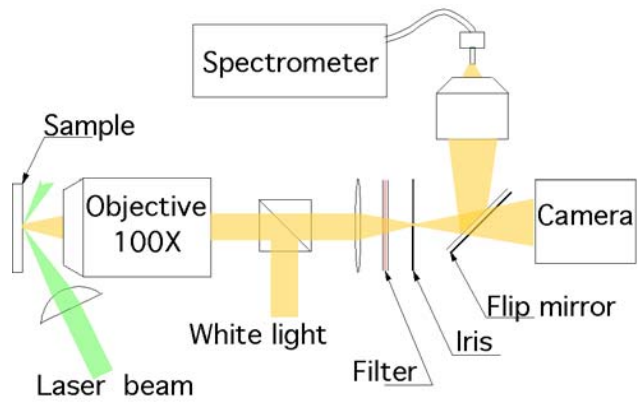

**FIGURE 4**

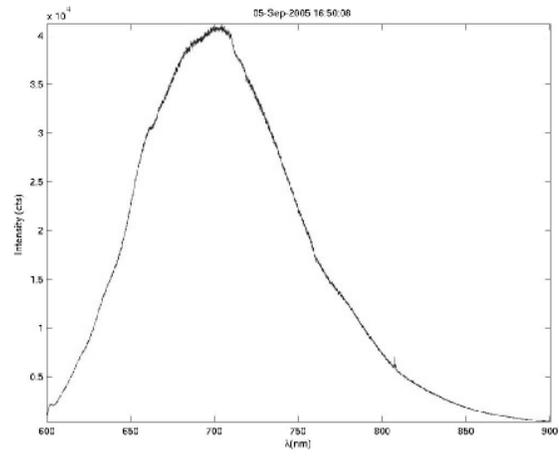

**FIGURE 5**